\documentclass{ifacconf}

\usepackage{graphicx}      
\usepackage{natbib}        
\begin{document}
\begin{frontmatter}

\title{Data Dissemination in Opportunistic Networks}


\author[First]{Radu I. Ciobanu} 
\author[First]{Ciprian Dobre} 

\address[First]{Computer Science Department, Faculty of Automatic Control and Computers, 
	University POLITEHNICA of Bucharest \\
	(e-mails: radu.ciobanu@cti.pub.ro, ciprian.dobre@cs.pub.ro)}

\begin{abstract}                
Mobile devices integrating wireless short-range communication technologies make possible new applications for spontaneous communication, interaction and collaboration. An interesting approach is to use collaboration to facilitate communication when mobile devices are not able to establish direct communication paths. Opportunistic networks, formed when mobile devices communicate with each other while users are in close proximity, can help applications still exchange data in such cases. In opportunistic networks routes are built dynamically, as each mobile device acts according to the store-carry-and-forward paradigm. Thus, contacts between mobile devices are seen as opportunities to move data towards destination. In such networks data dissemination is done using forwarding and is usually based on a publish/subscribe model. Opportunistic data dissemination also raises questions concerning user privacy and incentives. Such problems are addressed differently by various opportunistic data dissemination techniques. In this we analyze existing relevant work in the area of data dissemination in opportunistic networks. We present the categories of a proposed taxonomy that capture the capabilities of data dissemination techniques used in such networks. Moreover, we survey relevant data dissemination techniques and analyze them using the proposed taxonomy.
\end{abstract}


\end{frontmatter}

\section{Introduction}

Wireless short-range communication technologies (802.11b WiFi, Bluetooth, etc.) make possible a new and promising communication evolution called opportunistic networking. They are dynamically formed when mobile devices collaborate to form communication paths while users are in close proximity. In opportunistic networks mobile devices are data providers, data receivers, as well as data transmitters. They are based on the paradigm of store-carry-and-forward, as mobile devices act as data carriers to help disseminating the data, according to \cite{pelusi06}. Consider for example the case when user A wants to send a message to user B, but he/she has no network link to a wired access point, and cannot use long-range mobile telecommunication technologies (3G, WiMAX, etc.) because of the costs involved. Opportunistic communication is made possible by, for example, user C that arrives in the wireless transmission short-range of user A, receives the data, and further carries it towards user B.

Such spontaneous communication of mobile devices leads to the creation of opportunistic networks. Still such networks introduce problems such as how to decide if user C is the right carrier for the data, how to secure the data against malicious carriers, etc. An important topic in opportunistic networks is represented by data dissemination. In such networks, topologies are unstable. Various authors proposed different data-centric approaches for data dissemination, where data is proactively and cooperatively disseminated from sources towards possibly interested receivers, as sources and receivers might not be aware of each other, and never get in touch directly. Such data dissemination techniques are usually based on a publish/subscribe model. Opportunistic data-dissemination techniques were addressed by various authors. They suggested data dissemination techniques based on epidemic, social network, gossiping, or other algorithms. Still, there is no solution that can guarantee safe delivery, for example, of messages on a large-scale between drivers wanting to disseminate a specific traffic event.

In this we analyze existing work in the area of data dissemination in opportunistic networks. We analyze different collaboration-based communication solutions, emphasizing their capabilities to opportunistically disseminate data. We present the advantages and disadvantages of the analyzed solutions. Furthermore, we propose the categories of a taxonomy that captures the capabilities of data dissemination techniques used in opportunistic networks. Using the categories of the proposed taxonomy, we also present a critical analysis of four opportunistic data dissemination solutions. To our knowledge, a classification of data dissemination techniques has never been previously proposed.

The rest of the paper is structured as follows. Section 2 presents relevant contributions in the research area of opportunistic networks. Section 3 proposes the categories of a taxonomy for analyzing and comparing data dissemination techniques in opportunistic networks. In Section 4 we survey and critically analyze, using the proposed taxonomy, four relevant dissemination techniques. In Section 5 we conclude and present future research directions of our work.

\section{Related Work}

Opportunistic networking has been analyzed in many papers, but most of them treat forwarding, not dissemination. However, in recent years, several authors addressed the problem of data dissemination in opportunistic networks. Several taxonomies for forwarding algorithms have been proposed as well.

Authors of \cite{pelusi06} previously proposed a taxonomy for analyzing forwarding techniques. It separates them between algorithms without infrastructure (designed for completely flat ad-hoc networks) and algorithms with infrastructure (in which the ad-hoc networks exploit some form of infrastructure to opportunistically forward messages). Algorithms without infrastructure can be further divided into algorithms based on dissemination (like Epidemic, MV and Networking Coding), that are forms of controlled flooding, and algorithms based on context (like CAR and MobySpace), that use knowledge of the context that nodes are operating in to identify the best next hop at each forwarding step. Algorithms that exploit a form of infrastructure can also be divided into fixed infrastructure and mobile infrastructure algorithms. These algorithms have special nodes that are more powerful than the normal nodes. In case of fixed infrastructure algorithms (like Infostations and SWIM), special nodes are located at specific geographical points, whereas special nodes proposed in mobile infrastructure algorithms (like Ferries and DataMULEs) move around in the network randomly or follow predefined paths.

An alternative taxonomy, presented in \cite{conti09}, separates the forwarding methods according to their knowledge about context. Accordingly, there are three types of dissemination approaches: context-oblivious, partially context-aware and fully context-aware. The context-oblivious protocols do not exploit any contextual information about the behavior of users. The partially context-aware protocols exploit context information, but assume a specific model for this context. When the environment matches the assumed context, they perform very well, but their operation may not be correct if the environment is different from the assumption. Fully context-aware protocols learn and exploit the context around them and, while they may not be as efficient as partially context-aware protocols, they are much more adaptive. Some of the most popular forwarding algorithms nowadays are Bubble Rap, Propicman and HIBOp.

A thorough analysis of opportunistic networking is presented in \cite{conti10}. The authors present details regarding the architecture of Haggle and give various solutions to forwarding and data dissemination techniques. Also, security is discussed in terms of opportunistic networking, along with applications such as mobile social networking, sharing of user-generated content, pervasive sensing or pervasive healthcare.

Several papers exclusively treat the problem of data dissemination in opportunistic networks. The Epidemic approach is presented in \cite{vahdat00}. In \cite{yoneki07}, a dissemination technique based on publish/subscribe communication and communities is described, while \cite{lenders07} and \cite{lenders08} propose a wireless ad hoc podcasting system based on opportunistic networks. A multicast distribution method is presented in \cite{greifenberg08}, while ContentPlace, a system that exploits dynamically learned information about users' social relationships to decide where to place data objects in order to optimize content availability, is presented in \cite{boldrini10}. These methods are further analyzed in Section 4. To compare them we apply the categories of the proposed taxonomy. The obtained analysis highlights their advantages and disadvantages and further differentiates between their capabilities.

\section{A Taxonomy for Dissemination Techniques}

\begin{figure}[htp]
\begin{center}
\includegraphics[width=\linewidth]{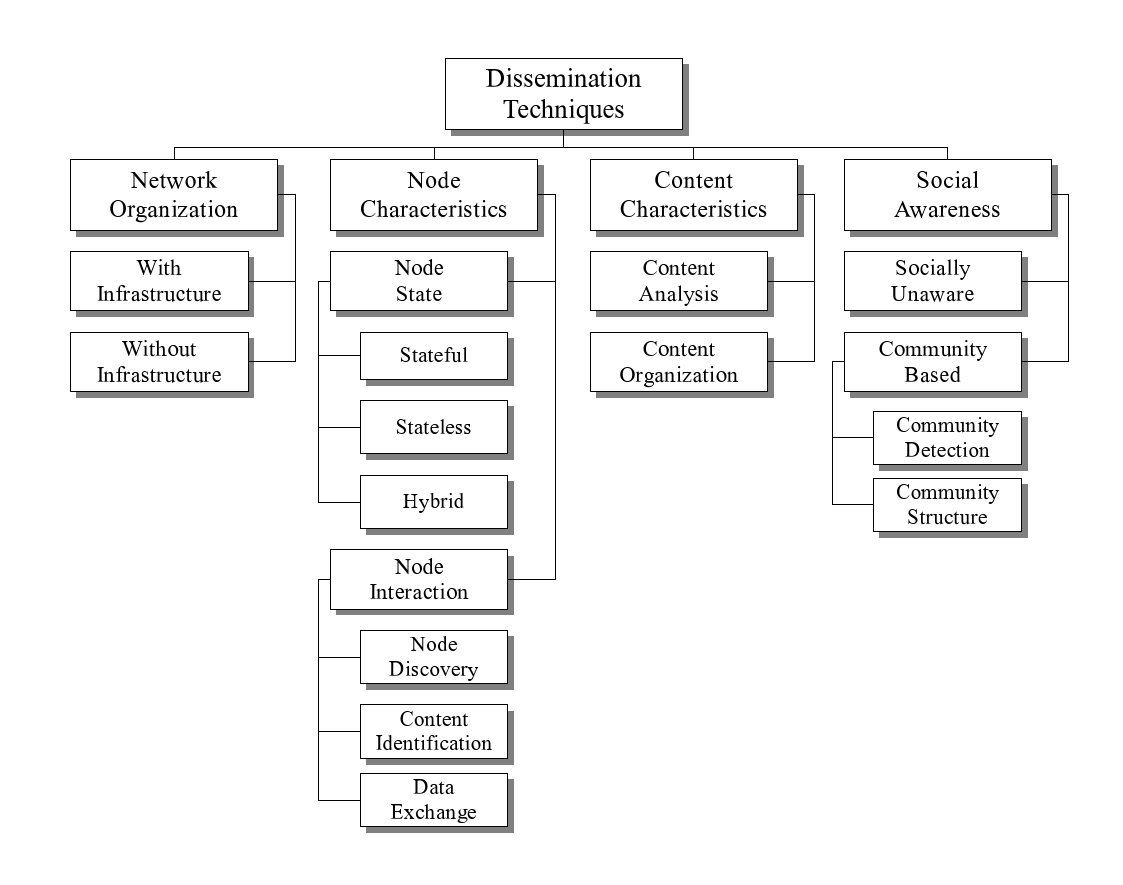}
\caption{A taxonomy for data dissemination techniques}
\label{fig:taxonomy}
\end{center}
\end{figure}

In this Section we introduce the categories of the proposed taxonomy for data dissemination techniques in opportunistic networks (Figure \ref{fig:taxonomy}).

According to the proposed taxonomy, data dissemination algorithms can be categorized by the \emph{organization of the network} on which they apply. In general, in an opportunistic network no assumption is made on the existence of a direct path between two nodes that wish to communicate. Nonetheless, some dissemination algorithms may exploit certain nodes called ``hubs'' and build an overlay network between them. The hubs are the nodes with the highest centrality in each community, where a node's centrality is the degree of its participation in the network. There are several types of node centrality relevant to data dissemination in opportunistic networks (such as degree centrality, betweenness centrality or closeness centrality), that will be detailed later. The algorithms that build an overlay network based on hubs fall under the category of data dissemination algorithms \emph{with infrastructure}. However, relying on an infrastructure might be costly to maintain (due to the large number of messages that have to be exchanged to keep it) and also highly unstable, especially in case of networks that contain nodes with a high degree of mobility. Considering this aspect, many data dissemination algorithms assume that the opportunistic network is a network \emph{without infrastructure}. The \emph{network organization} is relevant for a data dissemination algorithm because it directly influences the data transfer policies.

The actual nodes that participate in an opportunistic network play an important part in the way a data dissemination algorithm works. Consequently, the proposed taxonomy also categorizes dissemination techniques according to \emph{node characteristics}. A first characteristic of a node in an opportunistic network is the \emph{node state}. Depending on its implementation, a dissemination technique can either follow a \emph{stateful}, a \emph{stateless} or a \emph{hybrid} approach. An approach that maintains the state of a node requires control traffic (e.g. unsubscription messages in a publish/subscribe-based algorithm) that can prove to be expensive. Moreover, it suffers if frequent topology changes occur. On the other hand, a \emph{stateless} approach does not require control traffic, but has unsatisfactory results if event flooding is used. The \emph{hybrid} approach takes advantage of both the \emph{stateful} and the \emph{stateless} approaches.

Another important characteristic of a node in an opportunistic network is \emph{node interaction}. As stated before, nodes in an opportunistic network generally have a high degree of mobility, so the interaction between them must be as fast and as efficient as possible. The reason for this is that contact duration (the time interval for which two network devices can communicate when they come into range) may be extremely low. According to the proposed taxonomy, there are three basic aspects of \emph{node interaction}, the first one being \emph{node discovery}. Depending on the type of mobile device being used, the discovery of nodes that are in the wireless communication range can be done in several ways, but it is usually accomplished by sending periodical broadcast messages to inform neighboring nodes about the device's presence. When two nodes come into wireless communication range and make contact, they each have to inform the other node of the data they store. Therefore the second aspect of \emph{node interaction} is \emph{content identification}, meaning the way in which nodes represent the data internally and how they ``declare'' it (usually using some form of meta-data descriptions). Nodes may also advertise the channels they have data from or they may present a hash of the data they store. The final subcategory of \emph{node interaction} is \emph{data exchange}, which is the way two nodes transfer data to and from each other. This refers not only to the actual data transferring method, but also to the way data is organized or split into units. The three \emph{node interaction} steps presented here may also be done asynchronously for several neighboring nodes, and the way they are implemented affects the performance of a data dissemination algorithm.

As stated in \cite{conti10}, an interesting use case for opportunistic networks is the sharing of content available on mobile users' devices. In such a network, users themselves may generate content (e.g. photos, clips) on their mobile devices, which might be of interest to other users. However, content producers and consumers might not be connected to each other, so an opportunistic data dissemination method is necessary. Because there can be many types of content, each having different characteristics, the proposed taxonomy also classifies data dissemination algorithms according to the \emph{content characteristics}.

An important aspect of the actual content is its \emph{organization}. Most often, content is organized into channels, an approach used for publish/subscribe-based data dissemination. The publish/subscribe pattern is used mainly because communication is based on messages and can be anonymous, whilst participants are decoupled from time, space and flow. Time decoupling takes place because publishers and subscribers do not need to run at the same time, while space decoupling happens because a direct connection between nodes does not have to exist. Furthermore, no synchronized operations are needed for publishing and subscribing, so nodes are also decoupled from the communication flow. The approach allows the users to subscribe to a channel and automatically receive updates for the content they are interested in. Such an organization is taken further by structuring channels into episodes and enclosures.

Aside from the way content is organized at a node, the proposed taxonomy categorizes data dissemination techniques by \emph{content analysis}. \emph{Content analysis} represents the way in which the algorithm analyzes a certain content object and decides if it will fetch it or not. There are two reasons a node might download a content object from another encountered node: it is subscribed to a channel that the object belongs to, or the node has a higher probability of being or arriving in the proximity of another node that is subscribed to that channel, than the node that originally had the information. Not all dissemination algorithms analyze the data from other nodes: some simply fetch as much data as they can, until their cache is full, like Epidemic routing presented in \cite{vahdat00}, while others only verify if they do not already contain the data or if they have not contained it recently. More advanced \emph{content analysis} can be accomplished by assigning priorities (or utilities) to each content object from a neighboring node. In this way, considering the amount of free cache memory, a node can decide what and how many content objects it can fetch from another node. A node can also calculate the priority for its own content objects, and advertise only the priorities. Thus, a neighboring node can choose the data that maximizes the local priority of its cache. One method of computing priorities is based on heuristics that compare two content objects. Heuristics can compare content objects by their age, by their hop count or by the number of subscribers to the channel the object belongs to. A more complex approach to computing the value of priorities is to use a mathematical formula that assigns weights to various parameters. This method is used especially in socially-aware dissemination algorithms, where users are split in communities, and each community is assigned an individual weight (more about socially-aware algorithms will be presented in the next paragraph).

The final category of the proposed taxonomy is the \emph{social awareness}. Recently, the social aspect of opportunistic networking has been studied, because the actual nodes in an opportunistic network are represented by humans. They are the carriers of the mobile devices, so the human factor is an important dimension that must be considered by data dissemination algorithms. When designing such an algorithm, it is important to know that user movements are conditioned by social relationships. The first subcategory of \emph{social awareness} is represented by \emph{socially-unaware} algorithms, which do not assume the existence of a social structure that governs the movement or interaction of the nodes in an opportunistic network. Data dissemination techniques of this type may be as simple as spreading the content to all encountered nodes, but they can also take advantage of non-social context information such as geographical location.

Most of the recent data dissemination techniques that are aware of the social aspect of an opportunistic network are \emph{community-based}. \emph{Community-based} dissemination algorithms assume that users can be grouped into communities, based on strong social relationships between users. Even though there are several proposed representations of social behavior, the caveman model is by far the one mostly used (\cite{wu02}). Users can belong to more communities (called ``home'' communities), but can also have social relationships outside of their home communities (in ``acquainted'' communities). Communities are usually bound to a geographical space (static social communities), but they may also be represented by a group of people who happen to be in the same place at the same time (e.g. at a conference - temporal communities). According to this model, users spend their time in the locations of their home communities, but also visit areas where acquainted communities are located. As previously stated, a utility function may be used to decide which content objects must be fetched when two nodes are in range of each other. In a \emph{community-based} approach, each community would be assigned a weight, and the utility of a data object would be computed according to the community its owner comes from and the community of the (potentially) interested nodes.

One step that has to be executed before designing a \emph{community-based} dissemination algorithm is the \emph{community detection}. There are several methods used for organizing nodes from an opportunistic network into communities. One way is to simply classify nodes based on the number of contacts and contact duration of a node pair according to a threshold value, while another approach would be to define \emph{k}-clique communities as unions of all \emph{k}-cliques that can be reached from each other through a series of adjacent \emph{k}-cliques, as proposed in \cite{yoneki07}.

The phase following the detection of existing communities is the design of a \emph{community structure}. All nodes in a community can be identical (from the perspective of behavior), but there are also situations where certain nodes are more important in the dissemination scheme. As previously described, some data dissemination algorithms use network overlays constructed using hubs or brokers (e.g. nodes with the highest centrality in a community). The advantage of such an approach is that only nodes having a high centrality transfer messages to other communities. When a node wants to send a content object, it transfers it to the hub (or to a node with a higher centrality, which has a better chance of reaching the hub). The hub then transfers the object to the hub of the destination's community, where it eventually reaches the desired destination. The structure of a community has a high relevance in classifying data dissemination techniques, because a well-structured community can speed up the dissemination process significantly.

\section{Critical Analysis of Dissemination Algorithms}

In this Section we analyze the properties of four techniques for disseminating data in an opportunistic network, using the categories of the proposed taxonomy. The presented study evaluates the most relevant recent work in data dissemination algorithms. We also apply the proposed taxonomy to analyze and differentiate between the presented data dissemination techniques.

Authors of \cite{yoneki07} present a publish/subscribe data dissemination solution that uses a \emph{Socio-Aware Overlay} created on top of user-centric detected communities. The second data dissemination solution, proposed in \cite{lenders07} and \cite{lenders08}, uses a \emph{Wireless Ad Hoc Podcasting} system based on opportunistic networks. The next method is called the \emph{DTN Pub/Sub Protocol (DPSP)} and is an efficient publish/subscribe-based multicast distribution method for opportunistic networks, proposed in \cite{greifenberg08}. The final analyzed system for data dissemination is \emph{ContentPlace}. Presented in \cite{boldrini10}, it is a system that exploits dynamically learned information about users' social relationships to decide where to place data objects in order to optimize content availability.

\subsection{Socio-Aware Overlay}

The Socio-Aware Overlay algorithm proposed in \cite{yoneki07} is a data dissemination technique that creates an overlay for an opportunistic network with publish/subscribe communication. The overlay is composed of nodes having high centrality values that have the best visibility in a community. The data dissemination technique assumes the existence of a network \emph{with infrastructure}. This infrastructure is built by creating an overlay made of representative nodes from each community, where communities are detected using two different algorithms. The dissemination of subscriptions is done, together with the community detection, during the \emph{node interaction} phase, through gossiping. The gossiping dissemination sends each message to a random group of nodes, so from a \emph{node state} point of view, the Socio-Aware algorithm takes a \emph{hybrid} approach.

In order to choose an appropriate hub (or broker) in a network, the algorithm uses a measurement unit called node centrality. There are three proposed node centrality solutions: degree centrality (the number of direct connections), betweenness centrality (number of connections between two non-adjacent nodes) and closeness centrality (shortest paths to other nodes). The Socio-Aware algorithm uses the closeness centrality, so that the chosen broker maintains a higher message delivery rate.

\emph{Node discovery} is performed through Bluetooth and WiFi devices, while there are two modes of \emph{node interaction}, namely unicast and direct. The former is similar to Epidemic routing, while the latter provides a more direct communication mechanism like WiFi access points. From the standpoint of \emph{content organization}, the Socio-Aware algorithm is based on a publish/subscribe approach. At the \emph{data exchange} phase, subscriptions and unsubscriptions with the destination of community broker nodes are exchanged, as well as a list of centrality values with a time stamp. When a broker node changes upon calculation of its closeness centrality, the subscription list is transferred from the old one to the new one. Then, an update is sent to all the brokers. During the gossiping stage, subscriptions are propagated towards the community's broker. When a publication reaches the broker, it is propagated to all other brokers, and then the broker checks its own subscription list. If there are members in its community that must receive the publication, the broker floods the community with the information.

The Socio-Aware algorithm is a socially-aware \emph{community-based} algorithm, that has its own \emph{community detection} method. This method assumes a \emph{community structure} that is based on a classification of the nodes in an opportunistic network, from the standpoint of another node. A first type of node is one from the same community, having a high number of contacts of long/stable durations. Another type of node is called a familiar stranger and has a high number of contacts with the current node, but the contact durations are short. There are also stranger nodes, where the contact duration is short and the number of contacts is low, and finally friend nodes, with few contacts, but high contact durations.

In order to construct an overlay for publish/subscribe systems, \emph{community detection} is performed in a decentralized fashion, because opportunistic networks do not have a fixed structure. Thus, each node must detect its own local community. The authors propose two algorithms for distributed community detection, named Simple and \emph{k}-clique. In order to detect its own local community, a node interacts with encountering devices and executes the detection algorithm. The detection algorithm is done in the \emph{data exchange} phase of the interaction between nodes. Each node accomplishes the \emph{content identification} by maintaining information about the encountered nodes and contact durations (represented as a map called the familiar set) and the local community detected so far. When two nodes meet, a \emph{data exchange} is done, with each node acquiring information about the other's familiar set and local community. Each node then updates its local community and familiar set values, according to the algorithm used. As more nodes are encountered over time, the shape of the local community may be modified.

\subsection{Wireless Ad Hoc Podcasting}

The Wireless Ad Hoc Podcasting system, presented in \cite{lenders07} and \cite{lenders08}, extends podcasting to ad-hoc domains. The purpose is the wireless ad-hoc delivery of content among mobile nodes. Assuming a network \emph{without infrastructure}, the wireless podcasting service enables the distribution of content using opportunistic contacts whenever podcasting devices are in wireless communication range. From the standpoint of \emph{content organization}, the Ad Hoc Podcasting service employs a publish/subscribe approach. Thus, it organizes content in channels, which allows the users to subscribe and automatically receive updates for the content they are interested in. However, the channels themselves are divided into episodes and enclosures. Furthermore, enclosures are also divided into chunks, which are transport-level small data units of a size that can typically be downloaded in an individual node encounter. The reason for this division is the need for improving efficiency in the case of small duration contacts. The chunks can be downloaded opportunistically from multiple peers, and they are further divided into pieces, which are the atomic transport units of the network.

For \emph{node interaction} when two nodes are within communication range they associate and start soliciting episodes from the channels they are subscribed to. Since data is not being pushed, the nodes have complete control over the content they carry and forward. \emph{Node discovery} is done by using broadcast beacons sent periodically by each node. \emph{Content identification} is performed to identify channels and episodes at the remote peer that the current node is subscribed to. Two nodes in range exchange a Bloom filter hash index that contains all channel IDs that each node offers. Then each node checks the peer's hash index for channels it is subscribed to. The \emph{data exchange} phase begins if one of the nodes has found a matching channel. In this case, it starts querying for episodes. In order to perform \emph{content analysis}, the Wireless Ad Hoc Podcasting system proposes three different types of queries, employed according to the channel policy: a node requests any random episodes that a remote peer offers, a node requests episodes from the peer that are newer than a given date starting with the newest episode, or a node requests any episodes that are newer than a given date starting with the oldest episode.

When two nodes meet, and neither has content from a channel the other is subscribed to, several solicitation strategies are employed (\cite{lenders07}). They are used to increase the probability of a node having content to share with other nodes in future encounters. The solicitation strategies proposed are Most Solicited, Least Solicited, Uniform, Inverse Proportional and No Caching. The Most Solicited strategy fetches entries from feeds that are the most popular. The Least Solicited strategy does the opposite, by favoring less popular feeds. The Uniform strategy treats all channels equally, by soliciting entries in a random fashion, and has the advantage of being easy to implement. The Inverse Proportional strategy maintains a history list and solicits a feed with a probability which is inverse proportional to its popularity. Finally, No Caching is more of a benchmark for other strategies than a strategy itself, and assumes that a device has no public cache at all and that it stores or distributes only content from the fields it is subscribed to. Experiments show that the Uniform strategy has the best overall performance, while Inverse Proportional is the best one in regards to fairness.


\subsection{DPSP}

Authors of \cite{greifenberg08} propose a probabilistic publish/subscribe-based multicast distribution infrastructure for opportunistic networks based on DTN (Delay Tolerant Networking). The protocol uses a push-based asynchronous distribution delivery model. The idea is that nodes in the opportunistic network replicate bundles to their neighbors in order to get the bundle delivered by multiple hops of store-carry-and-forward.

As its name states, DPSP has a \emph{content organization} based on a channel subscription system, where users subscribe to channels and senders publish content. Although from the \emph{network organization} standpoint, DPSP assumes \emph{no infrastructure}, the nodes in the network are divided into three categories: sources, sinks and other nodes. Sources are the nodes that send content (in the form of bundles of data) to channels, while sinks subscribe to channels and receive information from them. The rest of the nodes are not interested in specific bundles, but they store, carry and forward bundles and subscriptions.

The \emph{node interaction} phase has several steps. When two nodes meet, \emph{content identification} is performed through the exchange of subscription lists. An entry in a subscription list contains the channel's URI, the subscription's creation time, its lifetime, the number of hops from the original subscriber to the current node, and an identifier for the subscription. Then, each node builds a queue of bundles to forward to the peer, and uses a set of filters to select the best. The selected bundles are subsequently sorted according to their priorities, and the \emph{data exchange} stage is performed by sending the bundles one by one until the contact finishes or the queue becomes empty.

In this approach, a set of filters is used in order to select the best bundles in a queue. Because the DPSP protocol is \emph{socially-unaware}, the filters used do not consider the organization of users into communities. There are three filters that handle the \emph{content analysis} and that can be used in any combination: Known Subscription Filter, Hop Count Filter and Duplicate Filter. The Known Subscription Filter removes bundles nobody is interested in, the Hop Count Filter removes bundles that are too old, while the Duplicate Filter removes bundles that the peer has already received. \emph{Content analysis} is also performed when the remaining bundles from a queue are sorted according to their priorities. Four heuristics are used to assign priorities to bundles: Short Delay, Long Delay, Subscription Hop Count and Popularity. Short Delay prefers newer bundles, Long Delay prefers older bundles, Subscription Hop Count sorts bundles according to hop count, and the Popularity heuristic sorts bundles by the number of nodes subscribed to the bundle's channel. The authors noticed that the Short Delay heuristic performs better with respect to delivery rates than the other heuristics.

\subsection{ContentPlace}

ContentPlace, proposed in \cite{boldrini10}, deals with data dissemination in resource-constrained opportunistic networks, by making content available in regions where interested users are present, without overusing available resources. To optimize content availability, ContentPlace exploits learned information about users' social relationships to decide where to place user data. The design of ContentPlace is based on two assumptions: users can be grouped together logically, according to the type of content they are interested in, and their movement is driven by social relationships.

For performance issues, ContentPlace assumes a network \emph{without infrastructure}. When a node encounters another node it decides what information seen on the other node should be replicated locally. For this it uses a local replication policy. First, when two nodes are in range, they have to discover each other. The \emph{node discovery} is not specified, but since the nodes are mobile devices it is probably done by WiFi or Bluetooth periodic broadcasts. For \emph{content identification}, nodes advertise the set of channels the local user is subscribed to upon encountering another node. ContentPlace defines a utility function by means of which each node can associate a utility value to any data object. When a node encounters another peer, it selects the set of data objects that maximizes the local utility of its cache, without violating the considered resource constraints. Due to performance issues, when two nodes meet, they do not advertise all information about their data objects, but instead they exchange a summary of data objects in their caches. Finally, the \emph{data exchange} is accomplished when a user receives a data object it is subscribed to when it is found in an encountered node's cache.

\emph{Content organization} in ContentPlace is done through channels to which users can subscribe. It assumes that the channel of a data object is decided by the source of the object at generation time. Consequently, unsubscription messages are not necessary, so a \emph{stateless} approach is used for the nodes. ContentPlace is a socially-aware, \emph{community-based} data dissemination algorithm. To have a suitable representation of users' social behavior, an approach that is similar to the caveman model proposed in \cite{wu02} is used, that has a \emph{community structure} which assumes that users are grouped into home communities, while at the same time having relationships in acquainted communities. For \emph{content analysis} nodes compute a utility value for each data object. The utility is a weighted sum of one component for each community its user has relationships with. The utility component of a data object for a community is the product of the object's access probability from the community members, by its cost (which is a function of the object's availability in the community), divided by the object's size. \emph{Community detection}, like at the Socio-Aware Overlay, uses the algorithms described in \cite{hui07}.

By using weights based on the social aspect of opportunistic networking, ContentPlace offers the possibility of defining different policies. There are five policies defined in \cite{boldrini10}: Most Frequently Visited (MFV), Most Likely Next (MLN), Future (F), Present (P) and Uniform Social (US). MFV favors communities a user is most likely to get in touch with, while MLN favors communities a user will visit next. F is a combination between MLN and MFV, as it considers all the communities the user is in touch with. In the case of P, users do not favor other communities than the one they are in, while at US all the communities the users get in touch with have equal weights.

\subsection{Analysis Results}

\begin{figure}[htp]
\begin{center}
\includegraphics[width=\linewidth]{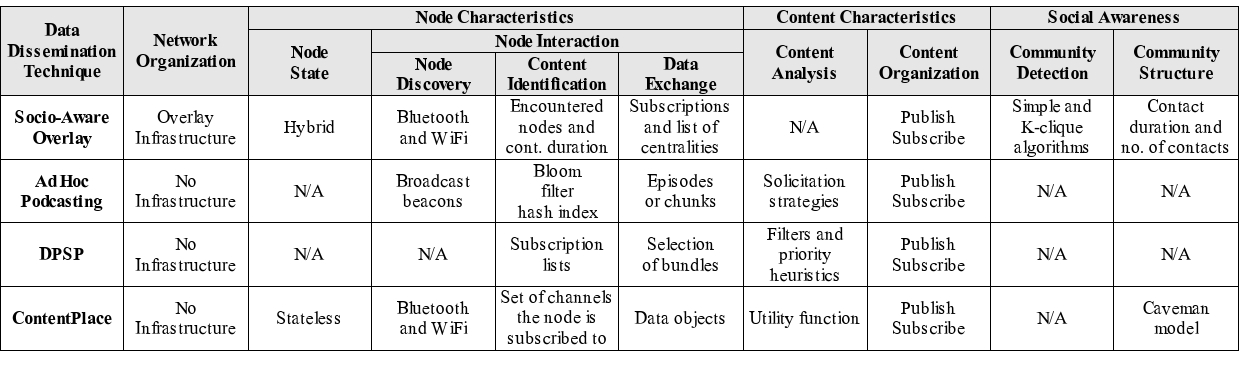}
\caption{Critical analysis of four dissemination techniques}
\label{fig:analysis}
\end{center}
\end{figure}

This Section presents a critical analysis of the four described protocols, according to the proposed taxonomy. The results of this analysis is presented in Figure \ref{fig:analysis}.

According to our analysis of the four solutions, only one assumes that the network over which data dissemination is performed has an infrastructure. The Socio-Aware Overlay algorithm builds an overlay infrastructure using the nodes with the highest centrality from each community. However, opportunistic networks generally contain nodes with a high degree of mobility, which make the task of creating and maintaining an infrastructure very hard to accomplish. The reason for this is that nodes may change communities very often (or they may not belong to a community at all), thus complicating the community detection phase. Furthermore, a device that is considered to be the central node (or hub) of a community may be turned off (due to different circumstances, like battery depletion), leaving the nodes in the hub's community without an opportunity to send messages to other communities, until a new hub is elected. Given these reasons, we believe that an approach that does not assume the existence of an infrastructure should be further considered.

The characteristics of a node from an opportunistic network play an important role in the structure of a data dissemination algorithm. Node characteristics refer to the way a node's state is represented and the way nodes interact when they are in contact. As stated in Section 3, the approach a data dissemination algorithm can take in regard to node state can be either stateless, stateful, or hybrid. Of the protocols we analyzed, ContentPlace chooses a stateless approach, while the Socio-Aware Overlay uses a hybrid representation of a node's state. The authors of the other two algorithms do not specify the node state, but we assume a stateful approach, because of the way the content is represented (for example, DPSP maintains subscription lists, for which node state is required). According to \cite{yoneki07}, a hybrid approach is the preferred solution because it takes advantage of both stateful and stateless approaches. Such an approach would not suffer under frequent topology changes, while at the same it would not require a large amount of control traffic.

The interaction between nodes has three steps that have been presented in detail in Section 3: node discovery, content identification and data exchange. Node discovery is usually done in the same way for all algorithms analyzed, but it may differ according to the type of devices that are present in the network. In case of the Socio-Aware Overlay and ContentPlace, the discovery is performed by using the Bluetooth or WiFi capabilities. The Ad Hoc Podcasting algorithm uses broadcast beacons, while the authors of DPSP do not mention a particular discovery method. It is a good approach to use the existing capabilities from the wireless protocols, but a data dissemination algorithm should try to extend the battery's life as much as possible. For example, when the battery is low, the broadcast beacons should be sent at larger time intervals.

Content identification, meaning the way in which nodes represent the data internally, also has a big impact in the efficiency of a data dissemination technique. The Socio-Aware Overlay maintains information about the encountered nodes and the duration of contacts, Ad Hoc Podcasting uses a Bloom filter hash index that contains all channel IDs, DPSP exchanges subscription lists and ContentPlace advertises the set of channels a node is subscribed to. The most efficient method is using Bloom filters, because they are space efficient data structures of fixed size that avoid unnecessary transmissions of data that the receiver has already received, according to \cite{bjurefors10}.

Data exchange should also be performed in a manner that optimizes the duration of a transfer. The nodes from the Socio-Aware Overlay exchange subscriptions and lists of centrality values, Ad Hoc Podcasting exchanges episodes or chunks, DPSP uses bundles and ContentPlace nodes exchange data objects. The smaller the data unit is, the bigger is the chance of a transmission to successfully finish, even in opportunistic networks where contact durations are very small. Therefore, one of the best approaches is the one employed by Ad Hoc Podcasting, where data is split into episodes and chunks.

The type of content organization that best suits opportunistic networks is the publish/subscribe pattern. The reason for this is that participants are decoupled from time, space and flow. Interested users simply subscribe to certain channels and receive data whenever the publishers post it. Publishers and subscribers do not have to be online at the same time, and it is not necessary that a direct connection exists between them. Consequently, all the analyzed data dissemination techniques organize their content according to a publish/subscribe approach. Content can also be analyzed in order for a node to decide what to download from an encountered peer. The Ad Hoc Podcasting technique uses five solicitation strategies that aim to increase the probability of a node having content to share with other nodes. DPSP has three filters used to select the best bundles in a queue and four heuristics that sort the remaining bundles. Finally, ContentPlace computes a utility function based on every community a node is in relationship with. The ContentPlace approach performs the best, because it takes advantage of the social aspect of opportunistic networking.

According to \cite{conti10}, human social structures are at the core of opportunistic networking. This is because humans carry the mobile devices, and it is human mobility that generates communication opportunities when two or more devices come into contact. Social-based forwarding and dissemination algorithms reduce by about an order of magnitude the overhead, compared to algorithms such as Epidemic routing. Therefore, the social aspect has a very important role in the efficiency of a data dissemination technique in an opportunistic network. Social awareness is based on the division of users into communities, which are defined as groups of interacting individuals organized around common values within a shared geographical location. Thus, an important step for socially-aware dissemination algorithms is community detection. Of the techniques we studied, only the Socio-Aware Overlay proposes its own community detection algorithms, called Simple and \emph{k}-clique. ContentPlace uses similar algorithms, while Ad Hoc Podcasting and DPSP are socially-unaware. As far as community structure goes, the Socio-Aware Overlay splits the nodes in a community from the standpoint of another node, according to the contact duration and number of contacts, while ContentPlace adopts a model similar to the caveman model. We consider that the future of data dissemination algorithms should be based on a socially-aware approach to take advantage of the human aspect of opportunistic networking.

After analyzing the four data dissemination techniques, we can conclude that there is no single best approach, but each algorithm provides certain aspects that offer advantages over the other implementations. In the next phase we plan to extend this work and propose a dissemination algorithm that uses the advantages of all analyzed solutions for maximum efficiency.

\section{Conclusions and Future Work}


In this we analyzed existing relevant work in the area of data dissemination in opportunistic networks. We presented the categories of a proposed taxonomy that capture the capabilities of data dissemination techniques used in such networks. Moreover, we critically analyzed four relevant data dissemination techniques using the proposed taxonomy. The purpose of the taxonomy, aside from classifying dissemination methods, has been to analyze and compare the strengths and weaknesses of the analyzed data dissemination techniques. Using this knowledge, we believe that an efficient data dissemination technique for opportunistic networks can be devised. We say that the future of opportunistic networking lies in the social property of mobile networks, so a great deal of importance should be given to this aspect.

In the future, we aim to propose and implement an opportunistic mobile wireless solution for communication based on the conclusions of our analysis. Such a solution can be used together with a context-aware platform for developing applications designed for mobile devices, with a focus towards recommendation and information of events towards users (such as the dissemination of academic events to all academic members). Such a solution might help in disseminating data between users having similar interests, even without the presence of dedicated wired access points and with lower costs than long-range mobile telecommunication protocols such as 3G or WiMAX. We believe that such a solution should be socially-aware, splitting nodes into communities (such as teachers, students, or students from the same group). An infrastructure may also be considered, built from nodes that are in contact with many communities (such as teachers or teaching assistants). Moreover, content should be exchanged between nodes based on the device owner's preferences, using context-aware data.


\begin{ack}
The research presented in this paper is supported by national project: ``TRANSYS - Models and Techniques for Traffic Optimizing in Urban Environments'', Contract No. 4/28.07.2010, Project CNCSIS-PN-II-RU-PD ID: 238. The work has been co-funded by the Sectoral Operational Programme Human Resources Development 2007-2013 of the Romanian Ministry of Labour, Family and Social Protection through the Financial Agreement POSDRU/89/1.5/S/62557. 
\end{ack}

\bibliography{ifacconf}             

\begin{thebibliography}{12}
\providecommand{\natexlab}[1]{#1}
\providecommand{\url}[1]{\texttt{#1}}
\providecommand{\urlprefix}{URL }
\expandafter\ifx\csname urlstyle\endcsname\relax
  \providecommand{\doi}[1]{doi:\discretionary{}{}{}#1}\else
  \providecommand{\doi}{doi:\discretionary{}{}{}\begingroup
  \urlstyle{rm}\Url}\fi

\bibitem[{Bjurefors et~al.(2010)Bjurefors, Gunningberg, Nordstrom, and
  Rohner}]{bjurefors10}
Bjurefors, F., Gunningberg, P., Nordstrom, E., and Rohner, C. (2010).
\newblock Interest dissemination in a searchable data-centric opportunistic
  network.
\newblock In \emph{Proc. European Wireless Conference}, EW 2010, 889--895.
  IEEE, Piscataway, NJ, USA.

\bibitem[{Boldrini et~al.(2010)Boldrini, Conti, and Passarella}]{boldrini10}
Boldrini, C., Conti, M., and Passarella, A. (2010).
\newblock {Design and performance evaluation of ContentPlace, a social-aware
  data dissemination system for opportunistic networks}.
\newblock \emph{Computer Networks}, 54, 589--604.

\bibitem[{Conti et~al.(2009)Conti, Crowcroft, Giordano, Hui, Nguyen, and
  Andrea}]{conti09}
Conti, M., Crowcroft, J., Giordano, S., Hui, P., Nguyen, H.A., and Andrea, P.
  (2009).
\newblock Routing issues in opportunistic networks.
\newblock In \emph{Middleware for Network Eccentric and Mobile Applications},
  121--147.

\bibitem[{Conti et~al.(2010)Conti, Giordano, May, and Passarella}]{conti10}
Conti, M., Giordano, S., May, M., and Passarella, A. (2010).
\newblock From opportunistic networks to opportunistic computing.
\newblock \emph{IEEE Communications Magazine}, 48, 126--139.

\bibitem[{Greifenberg and Kutscher(2008)}]{greifenberg08}
Greifenberg, J. and Kutscher, D. (2008).
\newblock Efficient publish/subscribe-based multicast for opportunistic
  networking with self-organized resource utilization.
\newblock In \emph{Proceedings of the 22nd International Conference on Advanced
  Information Networking and Applications - Workshops}, 1708--1714. IEEE
  Computer Society, Washington, DC, USA.

\bibitem[{Hui et~al.(2007)Hui, Yoneki, Chan, and Crowcroft}]{hui07}
Hui, P., Yoneki, E., Chan, S.Y., and Crowcroft, J. (2007).
\newblock Distributed community detection in delay tolerant networks.
\newblock In \emph{Proceedings of 2nd ACM/IEEE international workshop on
  Mobility in the evolving internet architecture}, MobiArch '07, 7:1--7:8. ACM,
  New York, NY, USA.

\bibitem[{Lenders et~al.(2007)Lenders, Karlsson, and May}]{lenders07}
Lenders, V., Karlsson, G., and May, M. (2007).
\newblock Wireless ad hoc podcasting.
\newblock \emph{2007 4th Annual IEEE Communications Society Conference on
  Sensor Mesh and Ad Hoc Communications and Networks}, 273--283.

\bibitem[{Lenders et~al.(2008)Lenders, May, Karlsson, and Wacha}]{lenders08}
Lenders, V., May, M., Karlsson, G., and Wacha, C. (2008).
\newblock Wireless ad hoc podcasting.
\newblock \emph{SIGMOBILE Mob. Comput. Commun. Rev.}, 12, 65--67.

\bibitem[{Pelusi et~al.(2006)Pelusi, Passarella, and Conti}]{pelusi06}
Pelusi, L., Passarella, A., and Conti, M. (2006).
\newblock Opportunistic networking: data forwarding in disconnected mobile ad
  hoc networks.
\newblock \emph{IEEE Communications Magazine}, 44, 134--141.

\bibitem[{Vahdat and Becker(2000)}]{vahdat00}
Vahdat, A. and Becker, D. (2000).
\newblock Epidemic routing for partially-connected ad hoc networks.
\newblock Technical report, {Duke University}.

\bibitem[{Wu and Watts(2002)}]{wu02}
Wu, J. and Watts, D.J. (2002).
\newblock Small worlds: the dynamics of networks between order and randomness.
\newblock \emph{SIGMOD Rec.}, 31, 74--75.

\bibitem[{Yoneki et~al.(2007)Yoneki, Hui, Chan, and Crowcroft}]{yoneki07}
Yoneki, E., Hui, P., Chan, S., and Crowcroft, J. (2007).
\newblock A socio-aware overlay for publish/subscribe communication in delay
  tolerant networks.
\newblock In \emph{Proceedings of the 10th ACM Symposium on Modeling, analysis,
  and simulation of wireless and mobile systems}, MSWiM '07, 225--234. ACM, New
  York, NY, USA.

\end{thebibliography}
\end{document}